\documentclass[a4paper,superscriptaddress,twocolumn,aps,pra,longbibliography]{revtex4-2}
\pdfoutput=1
\usepackage{amsmath,,amssymb}
\usepackage{tikz}
\usetikzlibrary{positioning}
\usepackage{graphicx}
\usepackage{braket}
\usepackage{algorithm, algorithmicx, algpseudocode, tabularx}
\usepackage{color}
\usepackage[normalem]{ulem}

\usepackage[utf8x]{inputenc}
\usepackage[colorlinks = true, citecolor=purple, linkcolor=blue, urlcolor=purple]{hyperref}

\makeatletter
\newcommand{\multiline}[1]{%
	\begin{tabularx}{\dimexpr\linewidth-\ALG@thistlm}[t]{@{}X@{}}
		#1
	\end{tabularx}
}
\makeatother

\let\oldReturn\Return
\renewcommand{\Return}{\State\oldReturn}

\newcommand{\cbox}[2]{\vcenter{\hbox{\includegraphics[width=#1em]{#2}}}}

\definecolor{darkgreen}{RGB}{0,100,0}
\definecolor{darkblue}{RGB}{0,0,255}

\begin{document}
	
	\title{Preentangling Quantum Algorithms - the Density Matrix Renormalization Group-assisted Quantum Canonical Transformation}
	
    \author{Mohsin Iqbal}
    \affiliation{Quantinuum, Leopoldstrasse 180, 80804 Munich, Germany}
    \author{David Muñoz Ramo}
    \affiliation{Quantinuum, 13-15 Hills Road, CB2 1NL Cambridge, United Kingdom}
    \author{Henrik Dreyer}
    \email{henrik.dreyer@quantinuum.com}
    \affiliation{Quantinuum, Leopoldstrasse 180, 80804 Munich, Germany}

	\begin{abstract}
	We propose the use of parameter-free preentanglers as initial states for quantum algorithms. We apply this idea to the electronic structure problem, combining a quantized version of the Canonical Transformation by Yanai and Chan [J. Chem. Phys. 124, 194106 (2006)] with the Complete Active Space Density Matrix Renormalization Group. This new ansatz allows to shift the computational burden between the quantum and the classical processor. In the vicinity of multi-reference points in the potential energy surfaces of H$_2$O, N$_2$, BeH$_2$ and the P4 system, we find this strategy to require significantly less parameters than corresponding generalized unitary coupled cluster circuits.
	We propose a new algorithm to prepare Matrix Product States based on the Linear Combination of Unitaries and compare it to the Sequential Unitary Algorithm proposed by Ran in [Phys. Rev. A 101, 032310 (2020)].
	\end{abstract}
	\maketitle
	
	Quantum computers beyond classical simulability have recently become available \cite{arute_quantum_2019}. How to proceed from the proof of concept stage to a useful application of quantum advantage remains an open problem. The key issue is that currently available qubits are too noisy to run deep quantum circuits as required for the most rewarding applications of quantum computers, e.g., quantum phase estimation. To bridge the gap between the noisy intermediate scale quantum (NISQ) era and fault-tolerant quantum computation, many researchers have turned to \emph{variational} quantum algorithms like the variational quantum eigensolver, the quantum approximate optimization algorithm, variational imaginary time evolution, variational quantum adiabatic algorithms and quantum neural networks \cite{peruzzo_variational_2014,schiffer_adiabatic_2021,bauer_hybrid_2016,nakanishi_sequential_2020,mcardle_error-mitigated_2019,cerezo_variational_2020, mcclean_theory_2016}. Variational quantum circuits are typically shallower and exhibit certain algorithmic resilience against noise. They work iteratively by evaluating the objective function of a given optimization problem on the quantum device and updating the variational parameters by using a classical optimization algorithm.
	
	Subsequently, it was realized that the corresponding optimization landscape is generically exceptionally hostile: circuit ansatzes that are sufficiently expressive suffer from gradients that vanish exponentially in the number of qubits, resulting in \emph{barren plateaus} that take exponential time to exit \cite{mcclean_barren_2018}. In the presence of noise, even less expressive, problem-inspired ansatzes suffer from barren plateaus, as long as a superlinear number of parameterized gates is to be trained \cite{wang_noise-induced_2021}. Reducing the number of variational parameters is therefore critical for the success of near-term variational quantum algorithms. Vanishing gradients also posed an early challenge to deep neural networks in classical machine learning. One of the key breakthroughs in this context was the \emph{pretraining}, i.e. the prior initialization of the learning parameters \cite{scholkopf_greedy_2007, hinton_fast_2006}. The concept of training a neural network  in a greedy, layer-wise fashion has recently been applied to the quantum setting \cite{skolik_layerwise_2021}.
	
	In this paper we initiate the study of an orthogonal approach: Instead of pretraining the parameters of a quantum circuit, we use classical resources to find \emph{parameter-free} quantum circuits, to which we append variational circuits. We hypothesize that, for problem sizes of practical use, this initialization places the quantum circuit close enough to the narrow gorge in the optimization landscape to ensure successful optimization.
	
	\begin{figure}[h!]
		\centering
		\fbox{\includegraphics[width=26em,scale=1]{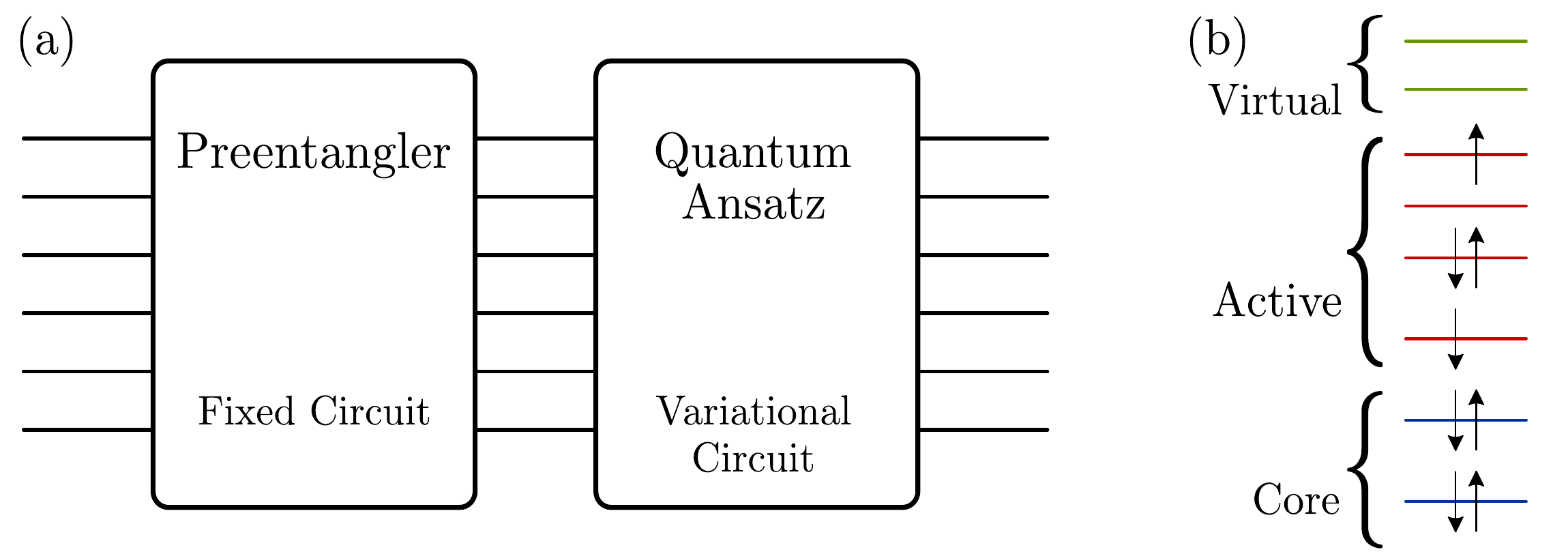}}
		\caption{ (a) Generic structure of a preentangled variational quantum algorithm. (b) Classification of spin orbitals in complete active space methods.}
		\label{fig:ct_circuit_meta}
	\end{figure}
	
	The resulting quantum circuits comprise two stages:  The \emph{preentangler}, a parameter-free circuit that can be optimized classically in an efficient and scalable manner; and the \emph{quantum ansatz}, a subsequent circuit of parametrised gates (Fig. \ref{fig:ct_circuit_meta}a). A good preentangler fulfills two criteria: (1) it lives in a  manifold of quantum states that is easy to optimize over classically and (2) shallow quantum circuits can be found for their preparation. In this paper, we focus on \emph{Matrix Product States} (MPS) as a preentangler and adapt the \emph{Canonical Transformation} (CT) \cite{yanai_canonical_2006} from quantum chemistry as the quantum ansatz. We apply this procedure to four problems in quantum chemistry with strong electronic correlations.
	
	The choice of preentangler and quantum ansatz is motivated by three facts. First, MPS can be obtained efficiently by the Density Matrix Renormalization Group (DMRG) algorithm. Even though DMRG is most efficient in the presence of  local one-dimensional interactions, the method has been shown to capture faithfully the Complete Active Space (CAS) of non-linear transition metal complexes \cite{doi:10.1063/1.2805383} and has been carried out for up to 100 orbitals.
	
	Second, significant work has been done on the conversion from MPS to the quantum circuits that prepare them. For example, Ran developed an iterative MPS preparation algorithm \cite{ran2020encoding}, based on the sequential generation with an ancilla \cite{schon2005sequential}. In this work we extend the body of MPS preparation algorithms by a novel strategy that uses the linear combination of unitaries that was introduced for doing Hamiltonian simulations on a quantum computer \cite{childs2012hamiltonian}.
	
	Finally, CT theory is a classical method designed for refining CAS-DMRG wavefunctions \cite{yanai_canonical_2006}. It is formulated in terms of unitary transformations. The hardness of implementing those classically can be overcome by approximations - or, as we propose here, by using a quantum computer.
	
	The combination of these three steps, classical CAS-DMRG, the preparation of the resulting MPS on a quantum computer, and the subsequent refinement using the canonical transformation on a quantum computer is what we call the Density Matrix Renormalization Group-assisted Quantum Canonical Transformation (DMRG-QCT).

	The outline of this paper is as follows. In Sec. \ref{sec:soa}, we introduce notation and CT theory. In Sec. \ref{sec:CT}, we formally introduce the DMRG-QCT method and investigate its potential to solve multi-reference problems with numerical experiments. In addition, we also analyze the resource requirements of DMRG-QCT. In Sec. \ref{sec:TSP}, we review the sequential unitary method and introduce the linear combination of the unitaries method for MPS preparation. We also compare the performance of the two methods. We conclude in Sec. \ref{sec:summary} with a summary and outlook.
	
	\section{Overview of Classical Canonical Transformation Theory}\label{sec:soa}
	Molecular electronic correlations are typically divided into two components. \emph{Static} correlation quantifies the part of electron correlation that is associated with multiple relevant determinants close in energy to the Highest Occupied Molecular Orbital (HOMO) and Lowest Unoccupied Molecular Orbital (LUMO). \emph{Dynamical} correlation describes corrections to the ground state arising from excitations involving low-lying core orbitals or high-lying virtual orbitals. When static correlation is negligible, we speak of \emph{single-reference} chemistry. In this situation, classical coupled cluster (CC) methods are generally sufficient to describe the dynamical correlation well (if done with sufficiently large basis sets) \cite{bartlett_applications_1994}. On the other hand, a \emph{multi-reference} situation occurs when a single determinant is not sufficient to describe the bonding, not even qualitatively. In the real world, these situation are found in chemical reactions (where valence configurations of products and reactants pass points of almost vanishing HOMO-LUMO gap),  excited states and  transition metal chemistry \cite{lyakh2012multireference}.
	
	To tackle multi-reference settings, one usually splits the molecular orbitals into \emph{core} ($c$), \emph{active} ($a$) and \emph{virtual} ($v$) orbitals (Fig \ref{fig:ct_circuit_meta}b), where the core orbitals are defined to be completely filled, orbitals in the active space are partially filled, and the orbitals in the virtual space are empty. The optimal way to assign an active space is unknown and initially relied on chemical intuition. Modern computer programs like BLOCK \cite{olivares-amaya_ab-initio_2015} and AutoCAS \cite{stein_automated_2016, stein_automated_2017, stein_delicate_2016, stein_measuring_2017} use the Fiedler vector as a proxy for the mutual information between orbitals, as well as machine learning algorithms to automate this process. The Hilbert space is then partitioned as $\mathcal{H} = \mathcal{H}_c \otimes \mathcal{H}_a \otimes \mathcal{H}_v$ (throughout the paper, we assume a Jordan-Wigner mapping from fermionic Fock space to qubits under which Slater determinants map to product states and $\ket{1}$ ($\ket{0}$) denotes occupied (unoccupied) orbitals).
	
    Many classical methods have been developed to deal with multi-reference situations, all of which are computationally expensive \cite{evangelista2018perspective}. For our quantum ansatz, we will adapt one of these methods, the Canonical Transformation (CT) Theory.
	
	The flavor of CT Theory that we use throughout the paper uses a CAS-DMRG ansatz
	\begin{align}
	\label{eq:ref_state}
	    \ket{\psi_0} = \ket{1}_c \otimes \ket{\text{MPS}}_a \otimes \ket{0}_v,
	\end{align}
	where
	\begin{align}
	    \ket{\text{MPS}}_a = \sum_{\substack{{i_1 \dots i_M} \\ \alpha_1 \dots \alpha_{M-1}}} A^{[1]}_{\alpha_1,i_1}  A^{[2]}_{\alpha_1 \alpha_2,i_2} \dots A^{[M]}_{\alpha_{M-1},i_M}  \ket{i_1 \dots i_M}.
	\end{align}
	Here, $M$ is the number of spin orbitals in the active space, physical indices $i_j$ run from $0$ to $1$ and $\alpha_j$ take values in $1 \dots D_j$, where $D_j$ is the \emph{bond dimension} of bond $j$. The bond dimension $D$ of the MPS is defined as the maximum over all $D_j$. The objects $A^{[j]}$ are matrices for $j \in \{1, M\}$ and three-legged tensors in the bulk. The cost of the optimisation scales as $\mathcal{O}(MD^3)$ and computations up to $D = 1000$ can be carried out routinely, while the largest reported bond dimension on a  supercomputer is $D=65536$ \cite{ganahl2022density}. In practice, much lower bond dimensions can yield accurate results, in particular for quasi one-dimensional geometries, where convergence has been achieved for active spaces of up to 100 orbitals \cite{baiardi_density_2020, doi:10.1021/acs.jctc.5b00174}.
	
	A major challenge of CAS-DMRG is to recover dynamical correlations. Single-reference CC methods are not applicable since the excitation operator assumes a fixed orbital occupancy of the reference state. One proposal to overcome this problem is Canonical Transformation (CT) Theory \cite{yanai_canonical_2006, yanai_canonical_2007, neuscamman_review_2010, yanai_canonical_2012, yanai_extended_2012}. In CT, the CC excitation operator is replaced by a unitary operator
	\begin{align}\label{eq:eT}
		U = e^T,
	\end{align}
	where
	\begin{align}
		T & = 
		\theta^{a_1}_{c_1} {\tau}^{a_1}_{c_1} + \theta^{a_1a_2}_{c_1c_2} {\tau}^{a_1a_2}_{c_1c_2} + \theta^{a_1a_2}_{a_3c_1} {\tau}^{a_1a_2}_{a_3c_1} \tag{core-active}	\\\nonumber\\
		& +\theta^{v_1}_{a_1} {\tau}^{v_1}_{a_1} + \theta^{v_1v_2}_{a_1a_2} {\tau}^{v_1v_2}_{a_1a_2} + \theta^{a_3v_1}_{a_1a_2} {\tau}^{a_3v_1}_{a_1a_2} \tag{active-virtual} 	\\\nonumber\\
		& +\theta^{a_2v_1}_{a_1c_1} {\tau}^{a_2v_1}_{a_1c_1} + \theta^{v_1v_2}_{a_1c_1} {\tau}^{v_1v_2}_{a_1c_1} + \theta^{a_1v_1}_{c_1c_2} {\tau}^{a_1v_1}_{c_1c_2} \tag{core-active-virtual} \\\nonumber
		& +\theta^{v_1}_{c_1} {\tau}^{v_1}_{c_1} + \theta^{v_1v_2}_{c_1c_2} {\tau}^{v_1v_2}_{c_1c_2}. \tag{core-virtual}
	\end{align}
	${\tau}^a_i = a^\dagger_i a_a - a^\dagger_a a_i$, and ${\tau}_{ij}^{ab} = a^\dagger_a a^\dagger_b a_i a_j - a^\dagger_i a^\dagger_j a_a a_b$. Indices labeled as $c_\bullet$, $a_\bullet$, and $v_\bullet$ correspond to orbitals in the core, active, and virtual spaces respectively and the repeated indices in each term are assumed to be summed over.
	
	The unitary is then applied to the reference state 
	\begin{align}
		\ket{\psi} = U \ket{\psi_0}.
	\end{align}
	The coefficients $\theta$ need to be optimized. Classically, this optimization proceeds as follows: The exponential can be expanded as
	\begin{align}
		\label{eq_energy}
		E &= \braket{\psi_0 | e^{-T} H e^{T} | \psi_0} \nonumber \\
		&= \braket{\psi_0 | H + [H,T] + \frac{1}{2} [[H,T],T] + \dots |\psi_0}.
	\end{align}
	On a classical computer, an exact, efficient treatment of (\ref{eq_energy}) is not possible. That is because expressions like $[[H,T],T]$ contain $N$-point correlation functions and, while 1-body and 2-body reduced density matrices  of the reference wavefunction can be computed, the computation of general $N$-body terms will necessarily take exponential time. In CT theory, higher-order terms are therefore approximated by sums of products of one- and two-body terms using the so-called \emph{cumulant} expansion \cite{shamasundar_cumulant_2009}. The convergence of the cumulant expansion is generally fast in single-reference scenarios but slow for  multi-reference systems. Ironically, these are precisely the systems that CAS-DMRG is particularly suited for \cite{hanauer_meaning_2012}. Thus, to remove the main limitation of CT Theory, alternative methods to compute~(\ref{eq_energy}) must be found.

	\section{Variational Quantum Eigensolver based on Canonical Transformation Theory}\label{sec:CT}
	
	To overcome the intractability of~\eqref{eq_energy}, we propose to variationally optimize~\eqref{eq_energy} on a quantum computer. We begin by describing the DMRG-QCT method before reporting our results on the capabilities of the method to tackle multi-reference problems and its resource requirements.

	\subsection{DMRG-QCT}
	\begin{figure}[t!]
		\centering
		\includegraphics[width=26em,scale=1]{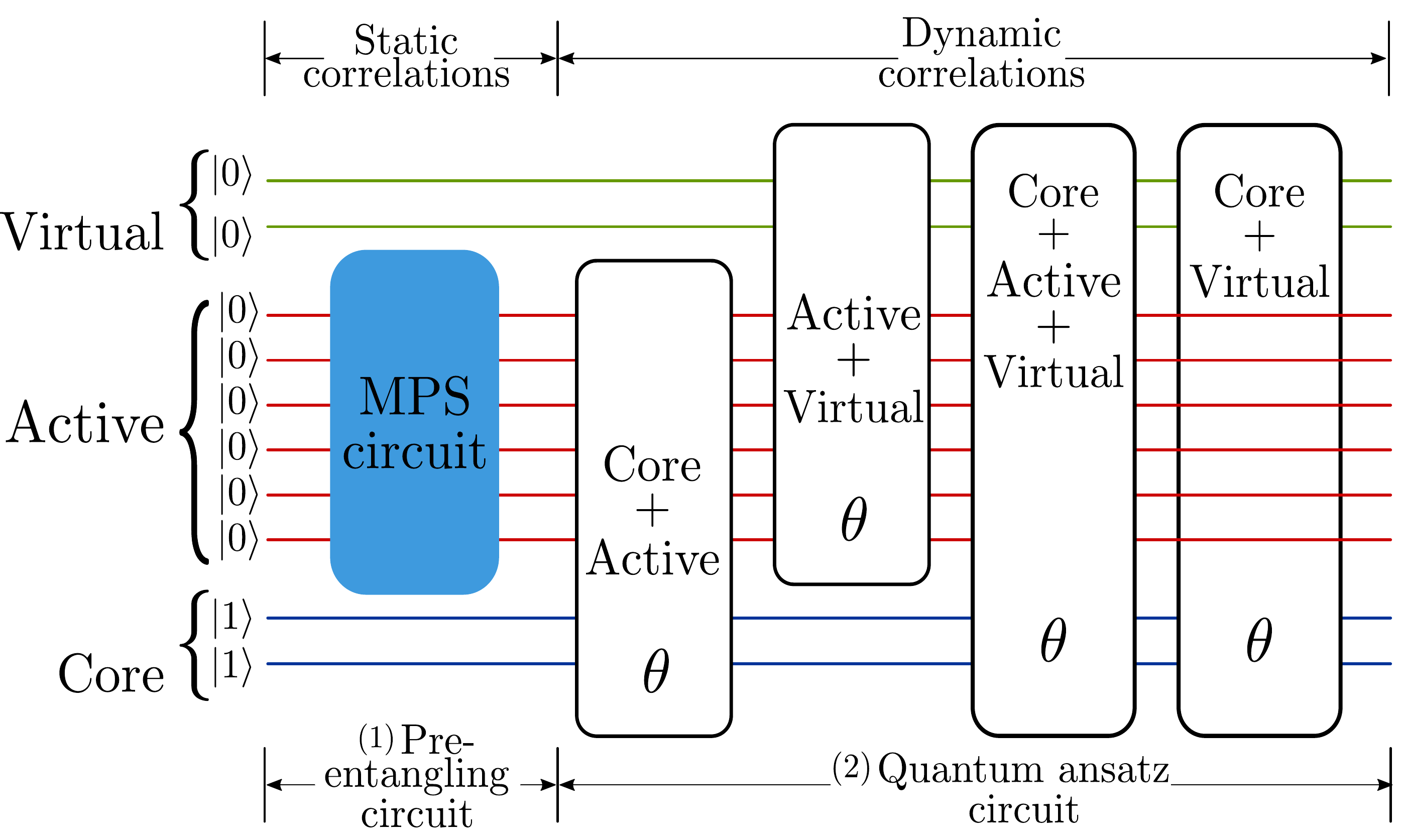}
		\caption{ Quantum circuit for DMRG-QCT. }	
		\label{fig:ct_circuit}
	\end{figure}
	DMRG-QCT uses DMRG to build static correlations in the wavefunction on a classical computer and then uses a CT inspired variational ansatz that we call Quantum Canonical Transformation (QCT) for adding dynamic correlations. As illustrated in Fig \ref{fig:ct_circuit}, the DMRG-QCT approach consists of two steps.
	
	In the first step, assuming that we have an adequate assignment of molecular orbitals to the active space, the algorithm approximates the ground state of the  Hamiltonian by the DMRG method. As an output from DMRG, we get the MPS description $\ket{\psi_0}$ of the ground state for the given bond dimension $D$. It is crucial for the unitary description to be reliable, i.e. the energy error of the unitary approximation should be sufficiently small, so as not to neglect too much of the static correlations in the MPS description. We discuss approaches to prepare MPS on the quantum computer in the next section.
	
	The second step is to construct a variational circuit to describe dynamic correlations by coupling active orbitals with orbitals in the core and virtual spaces. We can do this by implementing the quantum circuit for $e^T$ as defined in \eqref{eq:eT}. The exact circuit description of $e^T$ is intractable and we resort to an approximation, namely the first order Trotter-Suzuki decomposition \cite{hatano2005finding}.
	\begin{align}\label{eq:TSDecom}
		e^T \approx \prod_i e^{t_k \tau_k},
	\end{align}
	where for brevity, we write $T=\sum_k{t_k\tau_k}$, and the index $k$ runs over all the single and double excitations \emph{except} those which lie solely in the active space. Since the $\tau_k$s do not commute, the decomposition introduces a subtlety about the ordering of exponentials in \eqref{eq:TSDecom} and it could be consequential if the ordering is not chosen carefully \cite{grimsley2019trotterized}. Crucially, it has been pointed out in \cite{grimsley2019trotterized} that variation in the performance of different orderings is high in the presence of strong static correlations. For the DMRG-QCT method, however, the description of static correlations has already been taken care of by the active space DMRG computation, and so we expect a small variance in the performance for different ordering in the case of QCT. With the approximation of $U$ in \eqref{eq:TSDecom}, the goal of the DMRG-QCT method is to minimize the ground state energy,
	\begin{align}\label{eq:objective_func}
		E(\theta) = \bra{\psi_0} U^\dagger(\theta) H U(\theta) \ket{\psi_0},
	\end{align}
	by measuring it on the quantum computer and updating $\theta$ until we find a minimum.
	
	We can think of the DMRG-QCT method as an \emph{interpolation} between purely classical and purely quantum computation. By choosing the active space to contain all orbitals, we have a purely classical DMRG calculation. On the other hand, in the limit of an empty active space, we recover the fully quantum generalized unitary coupled cluster (GUCC) ansatz, a variant of the commonly used unitary coupled cluster ansatz that is suited for multi-reference computations.
	
	\subsection{Expressibility of DMRG-QCT}
	
	\begin{figure}[t!]
		\centering
		\includegraphics[scale=.9]{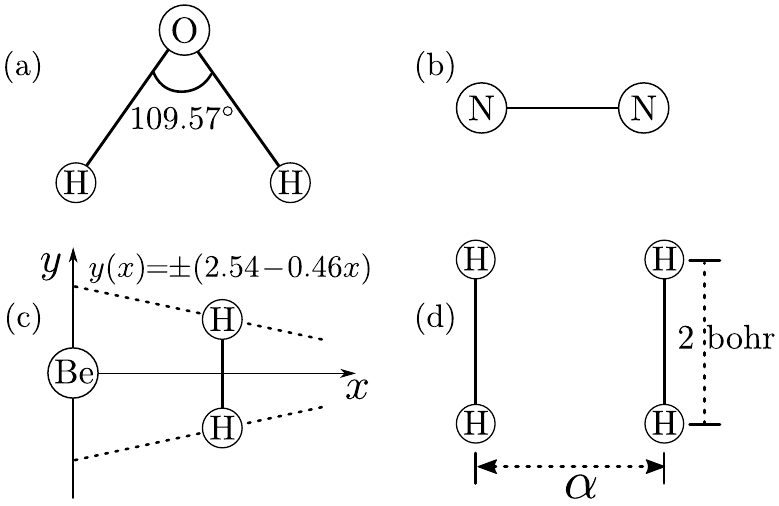}
		\caption{(a) Water molecule with an equilibrium bond angle. (b) Nitrogen dimer. (c) BeH$_2$ molecule. (d) P4 system consisting of two H$_2$ molecules.}
		\label{fig:molecules}
	\end{figure}
	
	\begin{figure*}[t!]
		\centering
		\includegraphics[width=52em,scale=1]{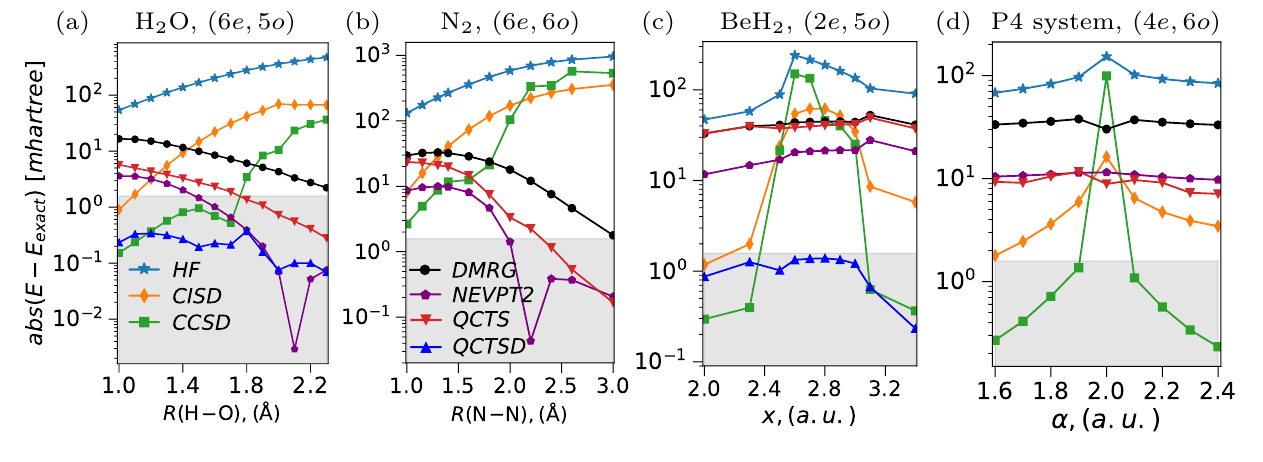}
		\caption{ Absolute energy error of different methods to approximate the ground state energies. Error is strictly positive for every method except for CCSD. (a) The water molecule in STO-6G basis set containing 14 qubits and the active space used for DMRG calculations contains $5 \times 2=10$ spin orbitals. (b) Nitrogen-dimer in STO-6G with an active space of 12 spin-orbitals. (c) BeH$_2$ in STO-6G basis set with an active space of 10 spin orbitals. (d) P4 model system with 6-31G basis set and the active space contains 12 spin orbitals. All the plots share the same legend that is shown in (a) and (b).}
		\label{fig:vqe}
	\end{figure*}
	
	To access the effectiveness of the DMRG-QCT ansatz method, we present our findings for four different molecular systems. These systems contain points in their potential energy surfaces (PES) which exhibit strong multi-reference behavior \cite{evangelista2006high, hanrath2008higher, lyakh2012multireference}. The PES of H$_2$O and N$_2$ describe typical bond-breaking situations. In the first problem, we consider the one-parameter dissociation of the water molecule at the equilibrium angle (Fig. \ref{fig:molecules}a) and study the ground state energy as the two hydrogen atoms are simultaneously pulled apart from the oxygen atom symmetrically. In the second example, we inspect the PES for the ground state of the Nitrogen molecule that is parametrised by the stretching of N-N bond distance (Fig. \ref{fig:molecules}b). In the third example, we investigate the transition state of the reaction between Be and H$_2$ \cite{purvis1983c2v}. The parametrization of the PES is expressed in Fig. \ref{fig:molecules}c. $x=0$ indicates the equilibrium situation with BeH$_2$ in linear configuration, and as the value of $x$ is increased, the system transitions into two non-interacting systems (Be and H$_2$) beyond the transition point. Finally, in the fourth example, we examine the P4 model system \cite{jankowski1980applicability} that consists of two hydrogen molecules (Fig. \ref{fig:molecules}d). The H-H bond distance in the two molecules is kept fixed, while one varies the distance between the molecules denoted by $\alpha$ in the PES. As one increases the value of $\alpha$, at $\alpha = 2$ bohr, the system obtains $D_{4h}$ symmetry. The higher symmetry leads to a configurational \emph{quasi-degeneracy}, and hence to the existence of strong static correlations in the wavefunction \cite{lyakh2012multireference}.
	
	We use the PySCF software package to compute one- and two-electron integrals \cite{sun2018pyscf}. For all problems, we use ChemMPS2 to perform the CAS-DMRG calculation \cite{CheMPS2cite1, CheMPS2cite2, CheMPS2cite3, CheMPS2cite4}. ChemMPS2 gives a symmetric MPS description of the ground state while exploiting SU(2), U(1) and point group symmetries. To accurately probe the potential of QCT variational circuit in expressing dynamic correlations, for the results here, we consider the exact DMRG construction of the ground state in the active space (i.e. the bond dimension $D$ used by DMRG is sufficiently large to represent the exact state), unless stated otherwise. Furthermore, once we have the MPS description of the ground state, we use the sequential unitary algorithm (see Sec. \ref{sec:TSP} for details) to obtain the circuit that prepares the MPS.
	
	Given a unitary construction of the ground state in the active space, we use InQuanto \cite{inquanto}, a platform that has been designed for doing computational chemistry problems on a quantum computer, to get the Jordan-Winger transformation of the fermionic Hamiltonian and implement the QCT ansatz. The realization of QCT involves the application of Trotter-Suzuki decomposition. To decompose $e^T$ as pointed out in \cite{grimsley2019trotterized}, we choose to order double excitations before the single excitations. We fix the arrangement of double excitations $\tau_{ij}^{ab}$ (labeled as $abij$) to be in the lexical order.  The exponential of the sum of the Pauli strings obtained from the Jordan-Wigner transformation of single ($\tau_{i}^{a}$) or double ($\tau_{ij}^{ab}$ ) excitations has been constructed by Pauli gadgets \cite{cowtan2019phase}.  We use pytket to decompose the DMRG-QCT circuit into elementary CNOT and one-qubit unitary gates~\cite{Sivarajah_2020}. The pytket compilation also performs certain non-trivial optimizations which lead to more efficient circuits.
	
	To optimize the energy, we iteratively update the variational parameters in the QCT circuit by using the  Broyden-Fletcher-Goldfarb-Shanno (BFGS) algorithm \cite{broyden1970convergence, fletcher1970new, goldfarb1970family, shanno1970conditioning}.
	
	The results of the DMRG-QCT method for the above mentioned molecules are given in Fig. \ref{fig:vqe}. We use the notation $(\bullet e,\bullet o)$ at the top of each plot to specify the number of electrons and orbitals in the active space used for DMRG calculations. We show data for Hartree-Fock (HF), configuration interaction with single and double excitations (CISD) and coupled cluster with single and double excitations (CCSD). We compute the energy error of different methods with respect to the full configuration interaction method in the given basis. As expected, the classical single-reference methods perform poorly. The error of CISD is substantial for stretched bonds (Fig. \ref{fig:vqe}(a,b)) and high symmetry points (Fig. \ref{fig:vqe}(c,d)). CCSD  becomes non-variational close to the multi-reference points.
	
	The data labeled as DMRG in Fig \ref{fig:vqe} shows the energy error in the absence of dynamic correlations; here, the variational QCT circuit acts trivially and static correlations generated by the preentangling MPS circuit completely determine the energy approximation. We also provide results for second order N-electron valence state perturbation theory (NEVPT2) applied on top of the multi-reference wavefunction from DMRG. Energy errors for DMRG-QCT with singles labeled as QCTS in Fig. \ref{fig:vqe} show the lowering of energies when the QCT circuit only contains variational parameters for the single excitations. It is worth pointing out that unitary coupled cluster ansatzes with only single excitations are typically not studied in the literature. This is due to Thouless Theorem \cite{thouless1960stability} which proves that single excitation clusters can not lower the energy of a single reference wavefunction.
	It is a distinct feature of QCTS that it manages to lower the energy by building dynamic correlations on top of the static MPS circuit.
	
	DMRG-QCT with single \emph{and} double excitations, labeled as QCTSD, is able to find a state with an error less than chemical accuracy in H$_2$O and BeH$_2$. In the case of BeH$_2$ and P4 system, NEVPT2 fails to achieve chemical accuracy.	We find \emph{the nonparallelism error} (NPE) \cite{lyakh2012multireference}, that is defined to be the absolute difference of maximum and minimum energy error, for NEVPT2 method to be $3.59\ mH$ (H$_2$O),  $9.93\ mH$ (N$_2$), $16.21\ mH$ (BeH$_2$),  and $1.75\ mH$ (P4 system). For the QCTSD method the NPE is $0.30\ mH$ (H$_2$O) and $1.15\ mH$ (BeH$_2$).
	
	Due to the computational and circuit requirements, which we discuss at length in the Sec \ref{sec:efficiency_DMRG-QCT}, we do not consider the DMRG-QCT ansatz with single and double excitations for N$_2$ and P4 system (Fig. \ref{fig:vqe}(b,d)). In contrast, the GUCCSD ansatz is computationally intractable for all the problems which we here consider. The results of QCTS and NVEPT2 are qualitatively similar for N$_2$ and P4 system.
	
\begin{figure}[t!]
		\centering
		\includegraphics[width=16em,scale=1]{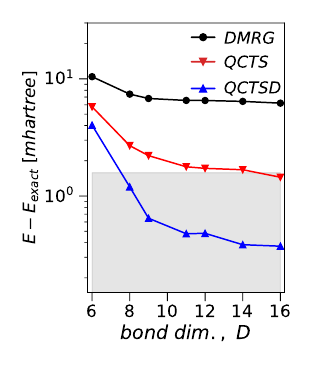}
		\caption{ Energy error as the function of bond dimension $D$ used in the DMRG algorithm for the water molecule at $R(\rm{O}\!-\!\rm{H})=1.8 \mbox{\normalfont\AA}$. $D=16$ is the bond dimension for the exact active space (i.e. without any approximation) as used Fig. \ref{fig:vqe}.}
		\label{fig:vqe_bond_dim}
	\end{figure}

	We have also examined the performance of the QCT ansatz with imperfect state construction by the DMRG method. We plot in Fig. \ref{fig:vqe_bond_dim} the effect on the performance of DMRG-QCT ansatz as we change the bond dimension of MPS describing the reference state $\ket{\psi_0}$ in \eqref{eq:ref_state}. It can be seen that both the QCTS and QCTSD show certain resilience against imperfect MPS construction by DMRG and the QCTSD ansatz maintains energy error below chemical accuracy. For a severe reduction in the bond dimension $D$ (which in the limit of $D \rightarrow 1$ corresponds to a single Hartree-Fock reference) QCT does not yield energies within chemical accuracy. At low bond dimension QCTS and QCTSD perform similarly, whereas for high bond dimension, QCTSD is very close to the exact energy. The data also suggests a preference for adding QCT single excitations over increasing the bond dimension utilized by the DMRG, e.g., the $D=6$ QCTS energy is already below the full active space energy.

	\subsection{Efficiency of DMRG-QCT}\label{sec:efficiency_DMRG-QCT}
	
\begin{figure}[t!]
		\centering
		\includegraphics[width=26em,scale=1]{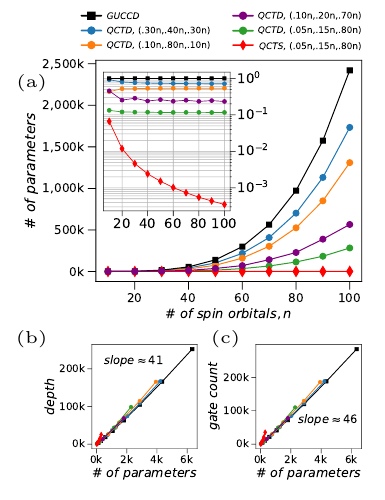}
		\caption{(a) the growth in the number of parameters for GUCC and DMRG-QCT ansatzes. The inset shows the same data, but normalized by the number of parameters in the GUCC ansatz. (b, c) the scaling of the circuit depth and number of gates as a function of variational parameters count. }
		\label{fig:excitations_cost}
	\end{figure}

	Now, we discuss the circuit complexity of implementing the DMRG-QCT ansatz and analyze it in comparison to the GUCC ansatz. We inspect the scaling of the number of variational parameters as a function of the number of spin orbitals or qubits $n$. We expect the number of variational parameters to be the most important factor in determining the cost (both the circuit requirements and time complexity) of variational quantum eigensolvers. A reduction in the number of parameters directly leads to a decrease in the depth and the elementary gate count.
	To corroborate this point, we show in Fig. \ref{fig:excitations_cost}(b, c) the cost of implementing an excitation/variational parameter in terms of circuit depth and elementary gates for different ansatzes and system sizes up to $n=36$. The large number of gates per variational parameter (46) implies that an ansatz with  few parameters is required for a variational algorithm to be run on a small-scall quantum computer for system sizes of practical use.
	Furthermore, the decrease in the number of parameters also reduces the the number of iterations required, i.e. the evaluation of energy by the quantum computer and the update of variational parameters by the classical optimizer. Estimation of the energy by the quantum computer is very expensive due to the enormous number of Pauli strings in the chemical Hamiltonian and so any reduction in the number of variational iterations leads to a substantial reduction in the runtime of the algorithm.

	\begin {table*}
	\begin{center}
		\bgroup
		\def\arraystretch{1.30}%
		\begin{tabular}{c @{\ }@{\ }|@{\ }@{\ } c @{\ }@{\ }|@{\ }@{\ } c @{\ }@{\ }|@{\ }@{\ } c @{\ }@{\ }|@{\ }@{\ } c @{\ }@{\ }|@{\ }@{\ } c @{\ }@{\ }|@{\ }@{\ } c @{\ }@{\ }|@{\ }@{\ } c} 
			Molecule & 
			\begin{tabular}{@{}c@{}} \# of qubits, \\ Active space \end{tabular} &
			\begin{tabular}{@{}c@{}} $D$, \\ $\mathcal{D}_{SEQ}$ \end{tabular} &
			\begin{tabular}{@{}c@{}} Ansatz \end{tabular}  &
			\begin{tabular}{@{}c@{}} \# of \\ parameters \end{tabular} &
			\begin{tabular}{@{}c@{}} CNOT\\ count \end{tabular} &
			\begin{tabular}{@{}c@{}} Gate\\ count \end{tabular} &
			\begin{tabular}{@{}c@{}} Depth \end{tabular}
			
			\\ \hline\hline
			
			\begin{tabular}{@{}c@{}}H$_2$O, \\ $R(\rm{O}\!-\!\rm{H})=1.7 \mbox{\normalfont\AA}$ \end{tabular} & 
			\begin{tabular}{@{}c@{}} 14 \\ STO-6G \\ $(6e,5o)$ \end{tabular} &
			\begin{tabular}{@{}c@{}} 22 \\ 20 \end{tabular} &
			\begin{tabular}{@{}c@{}} {\color{red}{QCTS}} \\ {\color{blue}{QCTSD}} \\ {\color{darkgray}{GUCCSD}} \end{tabular} & 
			\begin{tabular}{@{}r@{}} {\color{red}{20\ (0.03)}} \\ {\color{blue}{420\ (0.67)}} \\ {\color{darkgray}{623\quad}} \end{tabular} & 
			\begin{tabular}{@{}r@{}} {\color{red}{765\ (0.06)}} \\ {\color{blue}{9,905\ (0.75)}} \\ {\color{darkgray}{13,202\quad}} \end{tabular} & 
			\begin{tabular}{@{}r@{}} {\color{red}{2,050\ (0.09)}} \\ {\color{blue}{17,155\ (0.77)}} \\ {\color{darkgray}{22,360\quad}} \end{tabular} & 
			\begin{tabular}{@{}r@{}} {\color{red}{643\ (0.03)}} \\ {\color{blue}{13,590\ (0.72)}} \\ {\color{darkgray}{18,777\quad}} \end{tabular}
			\\ \hline
			
			\begin{tabular}{@{}c@{}}BeH$_2$, \\ $x=2.5\ (a.u.)$ \end{tabular} & 
			\begin{tabular}{@{}c@{}} 14 \\ STO-6G \\ $(2e,5o)$ \end{tabular} &
			\begin{tabular}{@{}c@{}} 8 \\ 32 \end{tabular} & 
			\begin{tabular}{@{}c@{}} {\color{red}{QCTS}} \\ {\color{blue}{QCTSD}} \\ {\color{darkgray}{GUCCSD}} \end{tabular} & 
			\begin{tabular}{@{}r@{}} {\color{red}{20\ (0.03)}} \\ {\color{blue}{420\ (0.67)}} \\ {\color{darkgray}{623\quad}} \end{tabular} & 
			\begin{tabular}{@{}r@{}} {\color{red}{847\ (0.06)}} \\ {\color{blue}{9,987\ (0.76)}} \\ {\color{darkgray}{13,202\quad}} \end{tabular} & 
			\begin{tabular}{@{}r@{}} {\color{red}{2,294\ (0.10)}} \\ {\color{blue}{17,399\ (0.78)}} \\ {\color{darkgray}{22,356\quad}} \end{tabular} & 
			\begin{tabular}{@{}r@{}} {\color{red}{745\ (0.04)}} \\ {\color{blue}{13,692\ (0.73)}} \\ {\color{darkgray}{18,777\quad}} \end{tabular}
			\\ \hline
			
			\begin{tabular}{@{}c@{}}P4 system, \\ $\alpha=1.9\ (a.u.)$ \end{tabular} & 
			\begin{tabular}{@{}c@{}} 16 \\ 6-31G \\ $(4e,6o)$ \end{tabular} & 
			\begin{tabular}{@{}c@{}} 29 \\ 90 \end{tabular} &
			\begin{tabular}{@{}c@{}} {\color{red}{QCTS}} \\ {\color{blue}{QCTSD}} \\ {\color{darkgray}{GUCCSD}} \end{tabular} & 
			\begin{tabular}{@{}r@{}} {\color{red}{24\ (0.02)}} \\ {\color{blue}{718\ (0.64)}} \\ {\color{darkgray}{1,120\quad}} \end{tabular} & 
			\begin{tabular}{@{}r@{}} {\color{red}{2,929\ (0.11)}} \\ {\color{blue}{20,189\ (0.78)}} \\ {\color{darkgray}{25,760\quad}} \end{tabular} & 
			\begin{tabular}{@{}r@{}} {\color{red}{8,443\ (0.20)}} \\ {\color{blue}{36,070\ (0.85)}} \\ {\color{darkgray}{42,375\quad}} \end{tabular} & 
			\begin{tabular}{@{}r@{}} {\color{red}{1,670\ (0.05)}} \\ {\color{blue}{25,546\ (0.71)}} \\ {\color{darkgray}{36,052\quad}} \end{tabular}
			\\ \hline
			
			\begin{tabular}{@{}c@{}}N$_2$, \\ $R(\rm{N}\!-\!\rm{N})=2.0 \mbox{\normalfont\AA}$  \end{tabular} & 
			\begin{tabular}{@{}c@{}} 20 \\ STO-6G\\ $(6e,6o)$ \end{tabular} & 
			\begin{tabular}{@{}c@{}} 32 \\ 210 \end{tabular} &
			\begin{tabular}{@{}c@{}} {\color{red}{QCTS}} \\ {\color{blue}{QCTSD}} \\ {\color{darkgray}{GUCCSD}} \end{tabular} & 
			\begin{tabular}{@{}r@{}} {\color{red}{48\ (0.02)}} \\ {\color{blue}{1,844\ (0.62)}} \\ {\color{darkgray}{2,955\quad}} \end{tabular} & 
			\begin{tabular}{@{}r@{}} {\color{red}{6,356\ (0.08)}} \\ {\color{blue}{55,836\ (0.71)}} \\ {\color{darkgray}{78,150\quad}} \end{tabular} & 
			\begin{tabular}{@{}r@{}} {\color{red}{18,136\ (0.15)}} \\ {\color{blue}{94,469\ (0.77)}} \\ {\color{darkgray}{122,405\quad}} \end{tabular} & 
			\begin{tabular}{@{}r@{}} {\color{red}{3,742\ (0.04)}} \\ {\color{blue}{70,579\ (0.66)}} \\ {\color{darkgray}{106,168\quad}} \end{tabular}
			\\ \hline
		\end{tabular}
		\egroup
	\end{center}
	\caption{Circuit requirements for the molecules considered. We show in parentheses ($\bullet$) the requirements of QCTS and QCTSD normalized with respect to GUCCSD. In the third column, we report the bond dimension $D$ of the MPS from DMRG, and the number of layers $\mathcal{D}_{SEQ}$ employed by the sequential unitary algorithm to approximate the MPS with the desired accuracy.}
	\label{table:molecules_cost}
	\end {table*}
	
	Let us now count the number of variational parameters required for different ansatzes. Each variational parameter corresponds to an excitation (single or double) in the first order Trotter-Suzuki approximation of $e^T$ in \eqref{eq:TSDecom}, and so the number of parameters is equal to the number of excitations. Furthermore, for the QCT ansatz, the number of electrons in the active space or the number of electrons in total does not change the number of excitations as it is also the case for GUCC ansatz. In Fig. \ref{fig:excitations_cost}a, we plot the scaling of variational parameters by counting the number of excitations for GUCC and DMRG-QCT ansatzes with respect to the number of spin orbitals. If $m$ is the number of orbitals (i.e. $m=n/2$), then for the GUCCD, the number of excitations is 
	\begin{equation}
	    {\frac{2}{3}}{m \choose 2} {m-2 \choose 2} + {{m \choose 2}}^2.
	\end{equation}
	The first (second) term counts for double excitations where the complete action of excitation is one (two) spin channels. The factor of $2/3$ in the first term accounts for the reduction obtained by enforcing the azimuthal spin symmetry. For the DMRG-QCT ansatz, we consider different splits $(c, a, v)$ by dividing $n$ spin orbitals into core $(c)$, active $(a)$, and virtual $(v)$ orbitals. As we increase the size of the active space, the number of variational parameters decreases (Fig. \ref{fig:excitations_cost}a(Inset)), e.g., for $a=0.8n$ we get a reduction by a factor of $2$. The reduction is more pronounced in a split with $c < a \ll v$ that represents a typical multi-reference calculation. For example, for the split $(0.05n, 0.15n, 0.8n)$, we get a reduction in the number of variational parameters by a factor of $10$.
	
	We also plot in Fig \ref{fig:excitations_cost}a, the scaling of QCTS. The growth in the number of parameters for QCTS is almost negligible compared to ansatzes with double excitations. Hence, the QCTS ansatz can be used to refine a DMRG wavefunction with the help of a quantum computer that is limited to a  small number of gates.
	
	We show in Table \ref{table:molecules_cost} resource requirements for the problems considered in Fig. \ref{fig:vqe}. Gate count and circuit depth for the QCT also include the cost for the MPS state preparation circuit, that is obtained by using the sequential unitary algorithm. Despite those extra costs, we observe a saving in all cost metrics of $20-40\%$ for QCTSD and $80-97\%$ for QCTS for the molecules considered. Note that the relative savings increase with system size, and for some entries in Table \ref{table:molecules_cost}, the gate count and depth is less than the scaling prefactor predictions in Fig. \ref{fig:excitations_cost}(b, c) because of the optimization performed by pytket during circuit synthesis.

	\section{Quantum Circuits to prepare Matrix Product States}\label{sec:TSP}
	In the preceding sections, we have assumed a perfect preparation strategy for MPS wavefunctions. In practice, short-depth quantum circuits can only prepare MPS wavefunction up to some error. Hence, we now describe different strategies to prepare MPS on a quantum computer. For the remainder of the section, we focus on MPS with physical dimension $d=2$ and denote the virtual bond dimension by $D$. We start by reviewing the sequential unitary (SEQ) algorithm to construct MPS. We discuss the effects of unitary freedom in the SEQ description of the MPS. Next, we introduce the linear combination of unitaries (LCU) algorithm as an alternative to the SEQ approach and compare their performance.
	
	\subsection{Sequential Unitary (SEQ) Algorithm}\label{subsec:SEQ}
	The SEQ algorithm was introduced by Ran in \cite{ran2020encoding}. It is motivated by the sequentially generated states ansatz which was first introduced for MPS by Sch{\"o}n et al. in \cite{schon2005sequential} and later generalized to Projected Entangled Pair States by Banuls et al. in \cite{banuls2008sequentially}.
	
	Here, we briefly discuss the key steps of the SEQ algorithm (see \cite{ran2020encoding} for details). The main ingredient of SEQ algorithm is the exact unitary construction of a MPS $|\psi[A^{[1]}, A^{[1]}, A^{[2]},
	\dots A^{[n]}]\rangle$, if $D=d$. Given such a MPS in canonical form \cite{perez2006matrix}, the algorithm proceeds by exploiting the isometric structure of  $A^{[j]}$ tensors and defines a unitary $G^{[j]}$ corresponding to each $A^{[j]}$,
	\begin{equation}
		\begin{aligned}[b]
			\cbox{3.1}{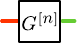}  &= \cbox{2.3}{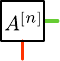} \\
			\cbox{3.35}{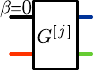}  &= \cbox{2.3}{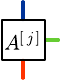}\ ,&\cbox{3.35}{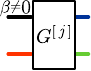}  &= \cbox{7.5}{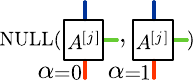} \\
			\cbox{3.35}{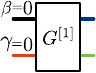} &= \cbox{2.3}{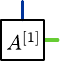}\ ,&  \cbox{3.35}{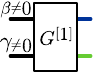} &= \cbox{5}{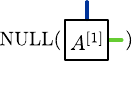},
		\end{aligned}
		\label{eq:SeqU}
	\end{equation}
	The second row defines unitaries corresponding to the intermediate tensors of the MPS. Each index has dimension 2 and the matching line colors on the two sides of each identity specifies the indices of every $G^{[j]}$. Every input of NULL($\bullet,\bullet,...$) is a vector in $d \times D=4$ dimensional space and its output represents an  orthonormal basis for the orthogonal complement of input vectors. The cascaded action of unitaries obtained by contracting the top-right index of $G^{[j]}$ with the bottom-left index of $G^{[j+1]}$ gives a global unitary $U$ and the of action of $U$ onto the product state gives the MPS, i.e., $|\psi\rangle = U|00\dots0\rangle$. 
	
	In the case of a generic MPS $|\psi\rangle$ with $D > d$, Ran realized that $U^\dagger$ acts as a disentangler. This insight leads to an iterative algorithm; it starts by initializing $|\psi_0\rangle=|\psi\rangle$ and each iteration consists of compressing the MPS to bond dimension $d$ and applying disentangling unitaries as follows,
	\begin{equation}
		\begin{aligned}[b]
			|\psi_i\rangle 
			\xrightarrow[\text{from } D \text{ to } d]{\text{Compress MPS}} \underset{U_i|00\dots \rangle}{\underset{=}{|\widetilde{\psi}_i\rangle}}
			\xrightarrow[\text{compress to } D_{max}]{\text{disentangle and}} \underset{ U_i^{\dagger}|\psi_i\rangle}{\underset{=}{|\psi_{i+1}\rangle}}
		\end{aligned}
		\label{eq:SeqU_arrows}
	\end{equation}
	Here we use $|\widetilde{\psi}\rangle$ to denote the truncated approximation of $|\psi\rangle$ to bond dimension $D=2$. The unitary $U_i$ is obtained as described earlier. The sequential application of unitaries is expected to transform $|\psi\rangle$ to a product state,  (i.e. $|00\dots0\rangle = \dots U_2^\dagger U_1^\dagger U_0^\dagger|\psi \rangle$) and the inverse action of $\mathcal{D}_{SEQ}$ layers of these unitaries
	\begin{equation}
		|\psi_{\mathcal{D}_{SEQ}}\rangle=U_0U_1\dots U_{\mathcal{D}_{SEQ}} |00\dots0\rangle
		\label{eq:seq_ansatz}
	\end{equation}
	gives the SEQ construction of a MPS. 
	
	\begin{figure}[t!]
		\centering
		\includegraphics[width=26.3em,scale=1]{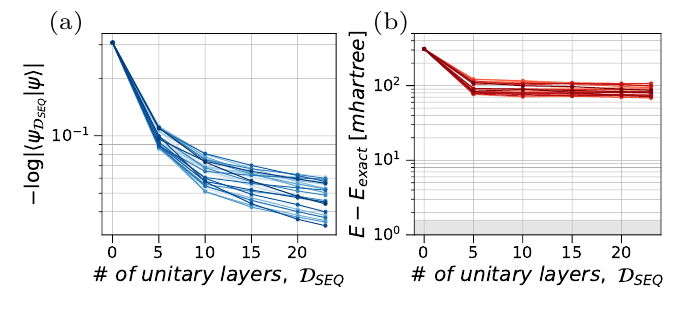}
		\caption{ (a) Fidelity of approximate state prepared by the SEQ algorithm with respect to the ground state of the N$_2$ molecule in a STO-6G basis set with 20 qubits. The atomic distance is $R(\rm{N}\!-\rm{N})=1.8$\AA. (b) shows the energy error for the same molecule. The gray region shows error below the chemical accuracy.}
		\label{fig:unitary_freedom}
	\end{figure}
	
	An important aspect of the SEQ algorithm is the freedom in the definition of local tensors. The choice of the orthonormal basis for the null spaces used in the definition of local unitaries $G^{[j]}$'s is not unique. We analyze the performance of SEQ algorithm with different choices of orthonormal basis at each unitary layer for the ground state $\ket{\psi}$ of the Nitrogen dimer (N$_2$). Fig. \ref{fig:unitary_freedom}(a,b) shows the convergence in fidelity and energy error for the unitary approximation $|\psi_{\mathcal{D}_{SEQ}}\rangle$ as the function of unitary layers. Distinct lines show the behavior of fidelity and error for arbitrary choices of null space basis in the local unitaries. Each of these trajectories corresponds to the particular choice of null space basis. We see that the choice of unitaries can improve the fidelity convergence, but not qualitatively. The crossing between different lines also suggests that the greedy optimization of unitaries at each layer would not necessarily lead to a globally optimal choice. Another way to exploit this unitary freedom is to use it in the gate synthesis (e.g. decomposition into one-qubit rotations and CNOT gates) of two qubit unitaries. By using isometric circuit techniques we can implement these state preparation unitaries (i.e. $G^{[j]}$'s) with savings in the number of CNOTs from $3$ to $2$ and one-qubit gates from $8$ to $6$ \cite{shende2006synthesis, iten2016quantum}.
	
	\subsection{Linear Combination of Unitaries (LCU) Algorithm}
	
	Now, we introduce a new method to prepare Matrix Product States. We approximate the MPS as,
	\begin{equation}
		\begin{aligned}[b]
			|\psi_{\mathcal{D}_{LCU}}\rangle = \left( U_0 + \kappa_1U_1 + \kappa_2U_2 + \dots\right)|00\dots0\rangle,
		\end{aligned}
		\label{eq:lcu}
	\end{equation}
	where each $U_i|00\dots0\rangle$ corresponds to a MPS with $D=2$ and the $\kappa_i$s are variational parameters.The generated state is a MPS that is block-diagonal, and we suspect that it captures better the symmetry structures of the target wavefunction. Moreover, in contrast to the product of unitaries, a generic linear combination of unitaries is not unitary, so in principle its correlations are not constrained by Lieb-Robinson bounds \cite{https://doi.org/10.48550/arxiv.2206.13527}.
	
	\begin{algorithm}[H]
		\caption{Find LCU approximation of $\ket{\psi}$}
		\label{algo:lcu}
		\begin{algorithmic}[1]
			\Procedure{LCU}{$\ket{\psi}, \mathcal{D}_{LCU}$} \Comment{$\mathcal{D}_{LCU}$ is the \# of $U_i$'s in \eqref{eq:lcu}}
			
			\State $\ket{{\psi}_0} \gets \text{Compress}(\ket{\psi},D=2)$
			\Comment{truncate MPS to $D$}
			\State $U_0 \gets \text{Unitary}(\ket{{\psi}_0})$

			\For{$i \gets 1$ to $\mathcal{D}_{LCU}$}  
			\State $ \ket{{\psi}_{i-1}}_{\text{proj}} \gets \text{Proj}(\ket{\psi}, \ket{{\psi}_{i-1}})$
			\Comment{projection of $\ket{\psi}$ }
			
			\State $\ket{{r}_i} \gets \text{Compress}(\ket{\psi} - \ket{{\psi}_{i-1}}_{\text{proj}},D=2)$
			
			\State $\kappa_i \gets \text{Optimize}(\ket{\psi}, \ket{{r}_i}, \ket{{\psi}_{i-1}})$
			
			\State $U_i \gets \text{Unitary}(\ket{{r}_i})$
			
			\State $\ket{{\psi}_i} \gets \ket{{\psi}_{i-1}} +\kappa_i \ket{{r}_i} $
			\State $\ket{{\psi}_i} \gets \text{Compress}(\ket{{\psi}_i},D=D_{max}) $
			
			\EndFor
			\Return{$\kappa$, $U$}
			
			\EndProcedure
		\end{algorithmic}
	\end{algorithm}
	
	The LCU algorithm works in an iterative fashion by estimating the residual between the target MPS and the current approximation (see Algorithm \ref{algo:lcu} for an implementation). The algorithm begins by compressing $\ket{\psi}$ to $D=2$ and initializing $\ket{\psi_0}$. We can find the exact unitary representation of $\ket{\psi_0}$ using \eqref{eq:SeqU} such that $\ket{\psi_0}=U_0\ket{00\dots0}$. At the start of each iteration, we compute the projection of the target state $\ket{\psi}$ on the current approximation $\ket{\psi_{i-1}}$, i.e.,
	\begin{align}
	    \ket{{\psi}_{i-1}}_{\text{proj}} = \frac{\langle  {\psi}_{i-1} | {\psi} \rangle}{\langle   {\psi}_{i-1} |  {\psi}_{i-1} \rangle} \ket{{\psi}_{i-1}}.
	\end{align}
	We compute the residual $\ket{r_i}$ by subtracting $\ket{{\psi}_{i-1}}_{\text{proj}}$ from $\ket{\psi}$ and compressing the resulting MPS to $D=d=2$. At this point it is important to note that one can also define a residual by taking the difference of $\ket{\psi}$ and $\ket{{\psi}_{i-1}}$ but this in general leads to a slowdown in the convergence of LCU algorithm. We find numerically that the residual obtained from $\ket{{\psi}_{i-1}}_{\text{proj}}$ performs better. Next, we optimize for the variational parameter $\kappa_i$, such that the normalized overlap between $\ket{\psi}$ and $\ket{\psi_{i-1}}+\kappa_i\ket{r_i}$ is maximized. This one parameter optimization is very fast given the fact that fidelities can be computed very quickly for MPS. Furthermore, without any loss of optimality, we can restrict the optimization to $\kappa_i>0$. Once we have $\kappa_i$, we can set a new $\ket{\psi_{i}}$ and if its bond dimension exceeds $D_{max}$ we compress it.
	
	\begin{figure}[t]
		\centering
		\includegraphics[width=24em,scale=1]{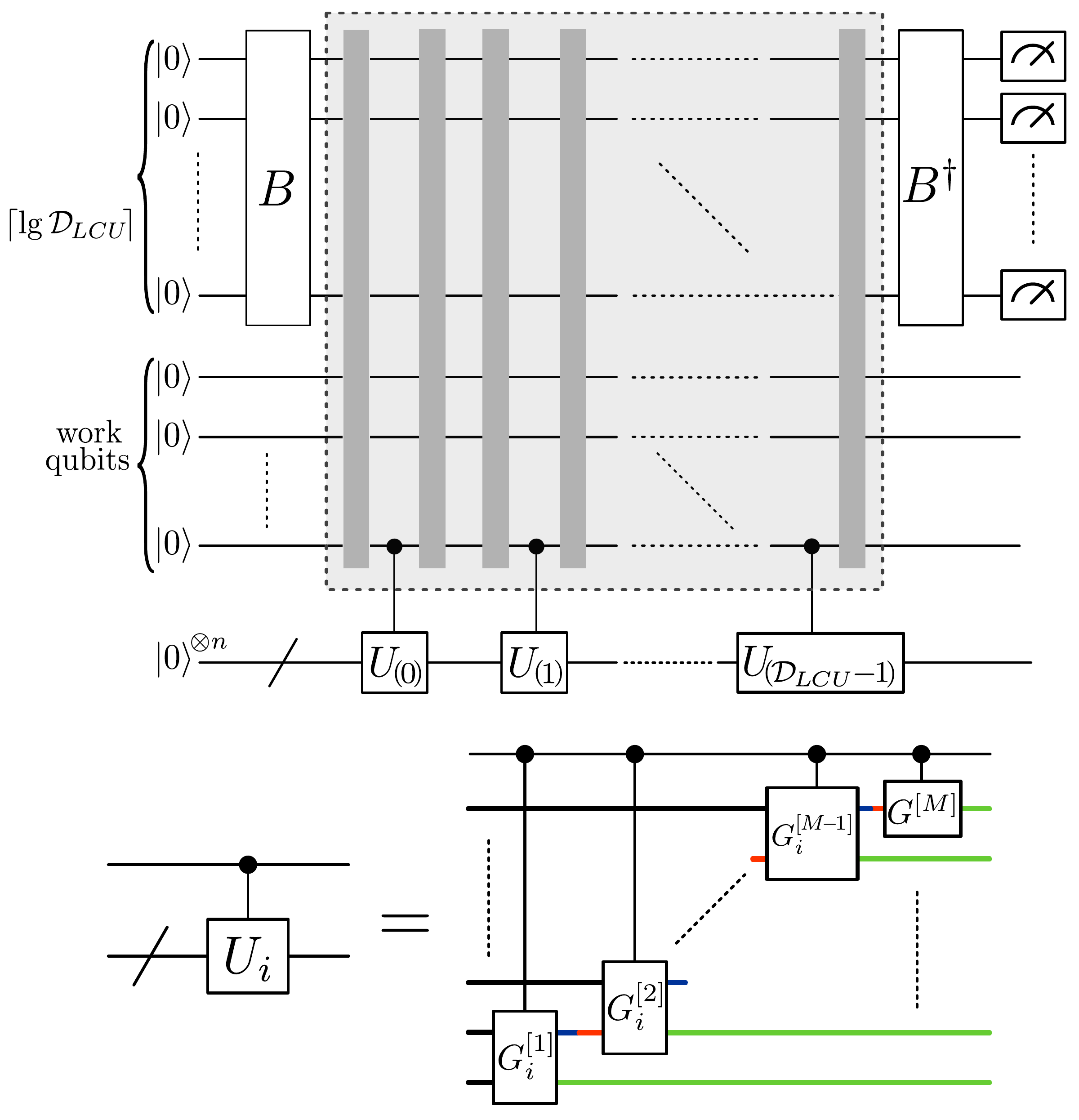}
		\caption{(Top) Non-deterministic circuit for constructing the linear combination of unitaries. (Bottom) Multi-qubit controlled unitary described in terms of one- and two-body gates as defined in \eqref{eq:SeqU}. }
		\label{fig:lcu_circuit}
	\end{figure}
	
	Now we consider the circuit implementation of the linear combination of unitaries. Since the linear combination of unitaries is not unitary, to achieve such an operation on a quantum device one needs to resort to a non-deterministic implementation. We have adapted the implementation given in \cite{childs2012hamiltonian},  employing the gray code construction \cite{rahman2014templates, babbush2018encoding, ralli2021implementation} (Fig. \ref{fig:lcu_circuit}). The box bordered by the dashed line in Fig. \ref{fig:lcu_circuit} indicates the implementation of $\lceil \lg \mathcal{D}_{LCU} \rceil$ controls for the unitaries which are ordered by their gray code. This ordering leads to a significant reduction in the number Toffoli and hence the CNOT and one-qubit gates. For the implementation of the multi-control unitary gates, there are two options. Naively, one could use a decomposition whose two qubit gate count scales quadratically. In order to avoid deep circuits, $\lceil \lg{\mathcal{D}_{LCU}} \rceil -1 $ further ancilla qubits can be added (called work qubits in Fig. \ref{fig:lcu_circuit})  \cite{barenco1995elementary}. Using this construction, the multi-control unitary gates can be implemented with a number of two qubit gates that is linear in the number of controls, and hence logarithmic in the number of layers $\mathcal{D}_{LCU}$ (which grows linearly with bond dimension). The first column of the multi-qubit gate $B$ is initialized with the coefficients $\kappa_i$. The rest of $B$ is set with an orthonormal subspace to make $B$ unitary. The circuit in Fig. \ref{fig:lcu_circuit} applies the correct unitary combination conditioned on the $\lceil \lg{\mathcal{D}_{LCU}} \rceil$ ancilla qubits being measured in the all-zeros state.
	
	\begin{figure}[t!]
		\centering
		\includegraphics[width=26em,scale=1]{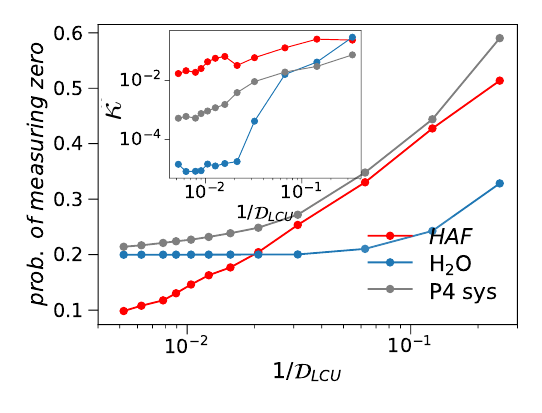}
		\caption{Probability of measuring zero for the LCU circuit as shown in Fig. \ref{fig:lcu_circuit}. Inset shows the data for convergence of $\kappa$’s as a function of $1/\mathcal{D}_{LCU}$. HAF labels data for the ground state of Heisenberg Antiferromagnet (HAF) chain, $H=\sum_i({{X_i X_{i+1}}+{Y_i Y_{i+1}}+{Z_i Z_{i+1}}})$ with $16$ sites. H$_2$O denotes data for the water molecule (see Fig. \ref{fig:molecules}a) in  STO-6G basis set with 14 qubits. P4 sys labels results for the ground state of P4 model system (see Fig. \ref{fig:molecules}d) in a 6-31G basis set with 16 qubits.}
		\label{fig:probability_conv}
	\end{figure}
	
	The cost of the LCU circuit in terms of elementary gates is dictated by the cost of realizing control two qubit unitaries. To implement control two qubit gates, we also exploit their isometric structure (i.e. only first columns of these unitaries are relevant, as discussed earlier in Sec. \ref{subsec:SEQ}). Furthermore, since only the first column of multi-qubit gates $B$ and $B^{\dagger}$ are relevant, they can be realized efficiently by using the state preparation strategy outlined in \cite{long2001efficient}. In the worst case $\mathcal{O}(\log{\mathcal{D}_{LCU}}\log{\log{\mathcal{D}_{LCU}}} )$ CNOT and one-qubit gates are required to implement $B$. While the overall cost of implementing the LCU circuit is linear in $\mathcal{D}_{LCU}$, the scaling prefactor is fairly large. For $n$ qubits, we need approximately 24$n\mathcal{D}_{LCU}$ CNOT and 31$n\mathcal{D}_{LUC}$ one-qubit gates. In contrast, to implement the SEQ circuit, the number CNOTS and one-qubit gates scales as 2$n\mathcal{D}_{SEQ}$ and 6$n\mathcal{D}_{SEQ}$ respectively. The better scaling of the SEQ circuit makes the SEQ algorithm more favorable in situations where it does achieve the desired accuracy.
	
		\begin{figure*}[t!]		
		\centering
		\includegraphics[width=52em,scale=1]{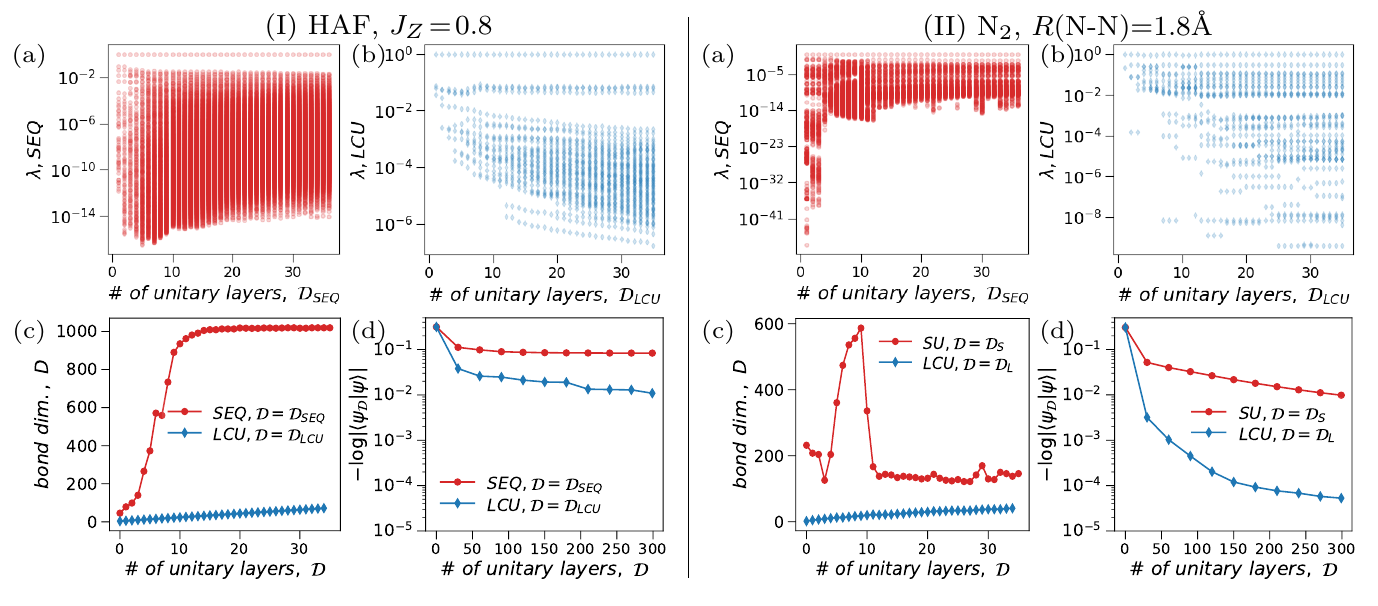}
		\caption{Left panel (I) shows results for the ground state of the Heisenberg Antiferromagnet (HAF) with $20$ spins. Right panel (II) shows data for the N$_2$
			molecule. (a,b) Schmidt values for the two algorithms without any truncation. (c) the bond dimension as a function of the number of layers. (d) Convergence of fidelity for the two algorithms with truncation to $D_{max}=3D_0$, where $D_0$ is the bond dimension of the target MPS.}
		\label{fig:schmidt_values}
	\end{figure*}
	
	We analyze the probability of measuring zero on ancilla qubits for different systems. We find the probability to converge to a constant value even for sufficiently large $\mathcal{D}_{LCU}$ (Fig. \ref{fig:probability_conv}). Although the convergence of probability can be attributed to the decay in LCU coefficient (Fig. \ref{fig:probability_conv}(Inset)), the fact that the limiting probability is fairly large for electronic structure problems makes the LCU algorithm especially useful to be applied for preparing the ground states of chemical Hamiltonians. Furthermore, this also alleviates the need for employing amplitude amplification or oblivious amplitude amplification that could incur considerable overhead \cite{grover1997quantum, berry2017exponential, guerreschi2019repeat}.

	\subsection{Numerical analysis of SEQ and LCU Algorithms}

	An important trait of the SEQ algorithm as also noted in \cite{ran2020encoding} is the existence of plateaus. Similar behavior has also been observed in a different context, which involved finding mixed state purifications \cite{hauschild2018finding}. In each iteration \eqref{eq:SeqU_arrows}, the application of disentangling unitaries raises the Schmidt rank and, hence, the bond dimension of the MPS. The latter increases exponentially with the number of layers $\mathcal{D}_{SEQ}$ and one must truncate it to some $D_{max}$ in order to keep the SEQ algorithm computationally viable. We analyze the behavior of Schmidt values (for the cut across maximum bond dimension) of the $|\psi_{i}\rangle$ for the SEQ algorithm without any truncation to $D_{max}$ (Fig. \ref{fig:schmidt_values}(I,II)a). The application of disentangling unitaries on the MPS introduces low weight Schmidt values, and this behavior holds for all the different models that we have studied. As a sidenote, the behavior of Schmidt values is quite distinct for the conventional many body in contrast to the chemical systems as can be seen in Fig. \ref{fig:schmidt_values}(I,II)a. Fig. \ref{fig:schmidt_values}(I,II)c shows the multiplicative increase in the bond dimension of the MPS before it saturates to a maximum value. Naively, one would expect to discard low weight Schmidt values (introduced by disentangling unitaries) without consequences. Unfortunately, this is not the case and the error introduced by compressing $|\psi_i\rangle$ to $D_{max}$ accumulates. This leads to the emergence of plateau as can be seen in Fig. \ref{fig:schmidt_values}(I,II)d. The slowdown of convergence is more pronounced in the energy error (see Fig. \ref{fig:unitary_freedom}b) where it flattens well before achieving chemical accuracy.
	
	In contrast, the LCU preparation strategy is an efficient classical algorithm even in the absence of truncation. During each iteration of the LCU algorithm, the bond dimension of the approximate MPS $\ket{\psi_i}$ increases in an additive manner (i.e. the bond dimension of $\ket{\psi_i}$ is $2i$). This is in contrast to the SEQ algorithm, where the bond dimension on the $i^{th}$ iteration without compression is expected to be $\mathcal{O}(2^i)$. Fig. \ref{fig:schmidt_values}(I,II)c shows the qualitative difference in the growth of bond dimension for the SEQ and LCU algorithms. In Fig. \ref{fig:schmidt_values}(I,II)b we show the Schmidt values of $\ket{\psi_{i}}$ for the LCU algorithm, which behave markedly different from those of the SEQ procedure. Furthermore, in Fig. \ref{fig:schmidt_values}(I,II)d we show that the LCU algorithm continues to increase fidelity with the target state even for large $\mathcal{D}$.
	
	\begin{figure*}[t!]
		\centering
		\includegraphics[width=48em,scale=1]{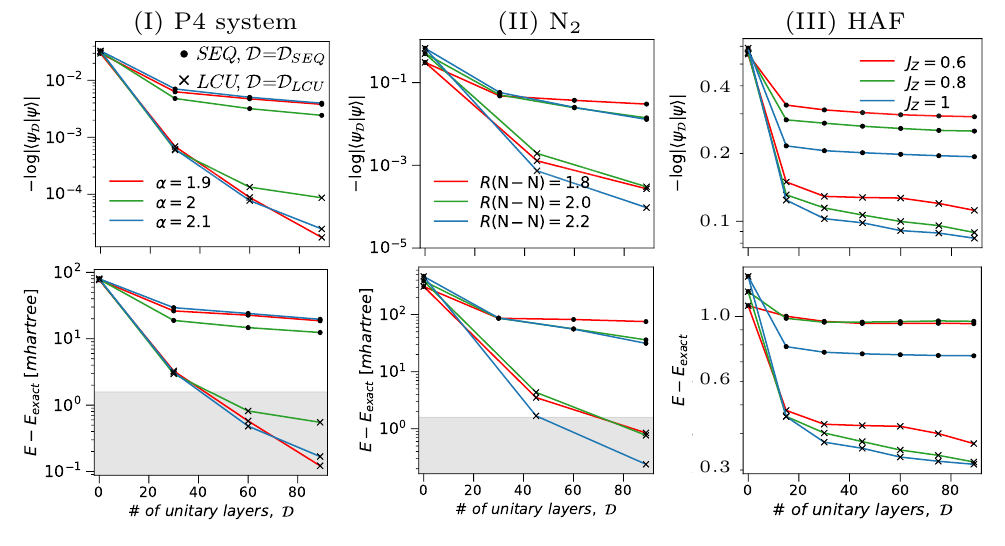}
		\caption{First (Second) row shows fidelity (energy error) of the SEQ and LCU preparation strategies with respect to the ground state. Panel (I) shows results for the P4 system in a 6-31G basis set with 16 qubits. In Panel (II) we show data for the N$_2$ molecule with 20 qubits. Panel (III) show results for the Heisenberg Antiferromagnet (HAF) with 32 spins.}
		\label{fig:tsp}
	\end{figure*}
	
	We now compare the performance of the SEQ and LCU algorithms for different chemical and spin systems. To use a given MPS preparation method in the context of chemical systems, as shown earlier in Sec. \ref{sec:CT} for the DMRG-QCT ansatz, it is crucial for the method to achieve error in the energy below chemical accuracy. That is because the energy error induced by the unitary approximation is a lower bound for the total energy error of the DMRG-QCT ansatz. In Fig. \ref{fig:tsp}, we present the behavior of the SEQ and LCU algorithm for the fidelity with true ground state and energy error. In the case of chemical systems (Fig. \ref{fig:tsp}(I,II)), we show data in the multi-reference regime while for the spin problem (Fig. \ref{fig:tsp}(III)), we show the behavior in the vicinity of the critical point. For a given number of layers, the LCU algorithm outperforms SEQ in all cases.

	\section{Summary and Outlook}\label{sec:summary}
	In this work, we have proposed a strategy to overcome some of the challenges of variational quantum algorithms. The strategy consists of two steps: first, a classically found trial wavefunction is prepared on a quantum computer and subsequently refined using short variational quantum circuits in a second step. We applied this idea in the context of quantum chemistry, where one natural choice is to assign the static and variational parts of the circuit to the static and dynamical correlations of the wavefunction, respectively. We found this strategy to lead to three kinds of advantages: First, existing ansatzes, e.g., GUCC ansatz can be modified to exclude the active space correlations which in our case leads to a system-size independent reduction of variational parameters by $30-90\%$, depending on the size of the active space.
	Second, a procedure to shift the computational burden between classical and quantum processors is implied, e.g., the computation can be tuned from a purely classical computation (where the full system is in the active space) to a purely quantum computation (where no active space precalculation is done).
	Third, and maybe most importantly, this strategy opens up the possibility of new, shallow ansatzes, that have previously not been considered. The availability of an entangled active space trial wavefunction allows the use of the QCTS ansatz, which for 100 spin orbitals contains roughly one thousand times less variational parameters than the GUCC ansatz.
	
	Several questions remain open: In this work, we have used a minimal basis for three of the four multi-reference problems. Since dynamic correlation typically is more important in larger basis sets, it would be useful to assess the dependence of the performance of QCTS and QCTSD on the size of the virtual space.
	
	While we have given one possible realization of the preentangler, the strategy is fairly flexible and can be utilized in a straightforward way to complement the capabilities of other techniques such as Adapt-VQE and symmetry preserving ansatzes \cite{grimsley2019adaptive, gard2020efficient}. Additionally, the preentangler approach can be used in the domains of quantum optimization and machine learning.
	
	One important step in the proposed approach is the approximate implementation of a MPS on the quantum computer. We have introduced a new algorithm for that purpose, the LCU algorithm. We have compared this algorithm to the SEQ algorithm from the literature and found it to be more expressive, albeit requiring a larger constant prefactor in its circuit decomposition. We have also found that the unitary freedom present in the canonical form of the MPS can be used both for the optimisation of the disentangler and for circuit synthesis, and conclude that the latter application is generally more powerful.
	
    It is also possible to prepare MPS by using a hybrid of SEQ and LCU procedures. We can think of two ways to combine the two methods. One idea is to modify \eqref{eq:seq_ansatz} to get an ansatz that is a sequence of operators where each operator itself is a linear combination of unitaries and hence not necessarily a unitary. Another option would be to consider a linear combination of sequential unitaries (i.e. each unitary in the sum is a product of unitaries) in \eqref{eq:lcu}. From these two propositions, the second approach can be implemented with minor changes in Algorithm~\ref{algo:lcu}. In the era of limited quantum resources, these generalized approaches could lead to better trade-offs between accuracy and circuit requirements for tensor state preparation.

	There are other ways for preparing MPS, e.g. it has been proposed in \cite{zhou2021automatically} to optimize unitary layers in the sequential unitary ansatz by automatic differentiation. We expect substantial improvements also for the LCU algorithm if one substitutes Algorithm~\ref{algo:lcu} with a similar unitary optimization procedure. Although that would incur increased classical computational overhead, it will result in higher-quality states within the variational manifold defined by \eqref{eq:lcu}. One can also consider preentanglers which consist of more generic tensor networks as given in \cite{Haghshenas_2022} via Quantum Circuit Tensor Networks. Preentanglers prepared by using an adiabatic algorithm~\cite{PhysRevResearch.4.023161} also seem promising. Moreover, the LCU algorithm described here can be used in other settings, e.g. for preparing low-energy states for a recently discovered quantum algorithm based on time series \cite{Lu_2021, https://doi.org/10.48550/arxiv.2206.01756}.

	Finally, while we have focused on the CT Theory, there exist other methods to recover dynamical correlation from the active space. The relative performance of, e.g. N-electron valence state second-order perturbation theory \cite{doi:10.1063/1.1361246} or adiabatic connection \cite{PhysRevLett.120.013001} deserves further investigation.
	
	\
	\begin{acknowledgments}
		We are indebted to Michal Krompiec for invaluable feedback, as well as David Zsolt Manrique for helpful discussions. This project was supported by the German Federal Ministry of Education and Research (BMBF) through the project EQUAHUMO (grant number 13N16069) within the funding program quantum technologies - from basic research to market.
	\end{acknowledgments}

	\bibliography{references}
	
	\onecolumngrid
	
	\begin{center}
		\rule{10cm}{0.5pt}
	\end{center}

\end{document}